\documentclass[twocolumn,trackchanges]{aastex62}
\pdfoutput=1 %for arXiv submission
\usepackage[T1]{fontenc}
\usepackage[figure,figure*]{hypcap}
\usepackage{amsmath,amstext,bm}
\usepackage{graphicx,textcomp,fancyhdr,hyperref}

\shorttitle{Obliquity of WASP-107\textnormal{b}}
\shortauthors{Rubenzahl et al.}
\newcommand{\istar}{i_\star}
\newcommand{\iorb}{i_\text{orb}}
\newcommand{\iorbb}{i_{\text{orb},b}}
\newcommand{\iorbc}{i_{\text{orb},c}}
\newcommand{\vsini}[0]{v\sin\istar}

\newcommand{\Vcbnn}{450} % it's actually negative
\newcommand{\Vcbnf}{250} % it's actually negative

\newcommand{\Mearth}{M_\oplus}
\newcommand{\Rearth}{R_\oplus}
\newcommand{\Mjup}{M_J}
\newcommand{\Rjup}{R_J}
\newcommand{\AU}{\text{AU}}

\begin{document}

\title{The TESS--Keck Survey IV: A Retrograde, Polar Orbit for the Ultra-Low-Density, Hot Super-Neptune WASP-107b}

% Authorlist
\author[0000-0003-3856-3143]{Ryan A. Rubenzahl}
\altaffiliation{NSF Graduate Research Fellow}
\affiliation{Department of Astronomy, California Institute of Technology, Pasadena, CA 91125, USA}

\author[0000-0002-8958-0683]{Fei Dai}
\affiliation{Division of Geological and Planetary Sciences, California Institute of Technology, Pasadena, CA 91125, USA}

\author[0000-0001-8638-0320]{Andrew W. Howard}
\affiliation{Department of Astronomy, California Institute of Technology, Pasadena, CA 91125, USA}

%%%%% Observers
%%%%% hires_j/j365.logsheet3
\author[0000-0003-1125-2564]{Ashley Chontos}
\altaffiliation{NSF Graduate Research Fellow}
\affiliation{Institute for Astronomy, University of Hawai`i, 2680 Woodlawn Drive, Honolulu, HI 96822, USA}

\author[0000-0002-8965-3969]{Steven Giacalone}
\affil{Department of Astronomy, University of California Berkeley, Berkeley, CA 94720, USA}

\author[0000-0001-8342-7736]{Jack Lubin}
\affiliation{Department of Physics \& Astronomy, University of California Irvine, Irvine, CA 92697, USA}

\author[0000-0001-8391-5182]{Lee J. Rosenthal}
\affiliation{Department of Astronomy, California Institute of Technology, Pasadena, CA 91125, USA}

\author[0000-0002-0531-1073]{Howard Isaacson}
\affiliation{Department of Astronomy, University of California Berkeley, Berkeley CA 94720, USA}
\affiliation{Centre for Astrophysics, University of Southern Queensland, Toowoomba, QLD, Australia}

%%%%% TKS Co-Is
\author[0000-0002-7030-9519]{Natalie M. Batalha}
\affiliation{Department of Astronomy and Astrophysics, University of California, Santa Cruz, CA 95060, USA}

\author{Ian J. M. Crossfield}
\affiliation{Department of Physics \& Astronomy, University of Kansas, 1082 Malott, 1251 Wescoe Hall Dr., Lawrence, KS 66045, USA}

\author[0000-0001-8189-0233]{Courtney Dressing}
\affiliation{Department of Astronomy, University of California Berkeley, Berkeley CA 94720, USA}

\author[0000-0003-3504-5316]{Benjamin Fulton}
\affiliation{NASA Exoplanet Science Institute/Caltech-IPAC, MC 314-6, 1200 E. California Blvd., Pasadena, CA 91125, USA}

\author[0000-0001-8832-4488]{Daniel Huber}
\affiliation{Institute for Astronomy, University of Hawai`i, 2680 Woodlawn Drive, Honolulu, HI 96822, USA}

\author[0000-0002-7084-0529]{Stephen R. Kane}
\affiliation{Department of Earth and Planetary Sciences, University of California, Riverside, CA 92521, USA}

\author[0000-0003-0967-2893]{Erik A Petigura}
\affiliation{Department of Physics \& Astronomy, University of California Los Angeles, Los Angeles, CA 90095, USA}

\author[0000-0003-0149-9678]{Paul Robertson}
\affiliation{Department of Physics \& Astronomy, University of California Irvine, Irvine, CA 92697, USA}

\author[0000-0001-8127-5775]{Arpita Roy}
\affiliation{Space Telescope Science Institute, 3700 San Martin Dr., Baltimore, MD 21218, USA}
\affiliation{Department of Physics and Astronomy, Johns Hopkins University, 3400 N Charles St, Baltimore, MD 21218, USA}

\author[0000-0002-3725-3058]{Lauren M. Weiss}
\affiliation{Institute for Astronomy, University of Hawai`i, 2680 Woodlawn Drive, Honolulu, HI 96822, USA}

%% Additional TKS team members (include based on TKS publication policy)
\author[0000-0001-7708-2364]{Corey Beard}
\affiliation{Department of Physics \& Astronomy, University of California Irvine, Irvine, CA 92697, USA}

\author{Michelle L. Hill}
\affiliation{Department of Earth and Planetary Sciences, University of California, Riverside, CA 92521, USA}

\author[0000-0002-7216-2135]{Andrew Mayo}
\affiliation{Department of Astronomy, University of California Berkeley, Berkeley CA 94720, USA}

\author[0000-0003-4603-556X]{Teo Mo\v{c}nik}
\affiliation{Gemini Observatory/NSF's NOIRLab, 670 N. A'ohoku Place, Hilo, HI 96720, USA}

\author[0000-0001-8898-8284]{Joseph M. Akana Murphy}
\altaffiliation{NSF Graduate Research Fellow}
\affiliation{Department of Astronomy and Astrophysics, University of California, Santa Cruz, CA 95064, USA}

\author[0000-0003-3623-7280]{Nicholas Scarsdale}
\affiliation{Department of Astronomy and Astrophysics, University of California, Santa Cruz, CA 95060, USA}

%%%%%%% Abstract
\begin{abstract}
We measured the Rossiter--McLaughlin effect of WASP-107b during a single transit with Keck/HIRES. We found the sky-projected inclination of WASP-107b's orbit, relative to its host star's rotation axis, to be $|\lambda| = {118}^{+38}_{-19}$~degrees. This confirms the misaligned/polar orbit that was previously suggested from spot-crossing events and adds WASP-107b to the growing population of hot Neptunes in polar orbits around cool stars. WASP-107b is also the fourth such planet to have a known distant planetary companion. We examined several dynamical pathways by which this companion could have induced such an obliquity in WASP-107b. We find that nodal precession and disk dispersal-driven tilting can both explain the current orbital geometry while Kozai--Lidov cycles are suppressed by general relativity. While each hypothesis requires a mutual inclination between the two planets, nodal precession requires a much larger angle which for WASP-107 is on the threshold of detectability with future Gaia astrometric data. As nodal precession has no stellar type dependence, but disk dispersal-driven tilting does, distinguishing between these two models is best done on the population level. Finding and characterizing more extrasolar systems like WASP-107 will additionally help distinguish whether the distribution of hot-Neptune obliquities is a dichotomy of aligned and polar orbits or if we are uniformly sampling obliquities during nodal precession cycles.
\end{abstract}

% Intro
\section{Introduction} \label{sec:intro}

WASP-107b is a close-in ($P = 5.72$~days) super-Neptune orbiting the cool K-dwarf WASP-107. Originally discovered via the transit method by WASP-South, WASP-107b was later observed by K2 in Campaign 10~\citep{Howell2014}. These transits revealed a radius close to that of Jupiter, $R_b = 10.8 \pm 0.34~\Rearth = 0.96 \pm 0.03~\Rjup$~\citep{DaiWinn17, Mocnik17, Piaulet2020}. However, follow-up radial velocity (RV) measurements with the CORALIE spectrograph demonstrated a mass of just $38 \pm 3~\Mearth$~\citep{Anderson2017}, meaning this Jupiter-sized planet has just one-tenth its density. Higher-precision RVs from Keck/High Resolution Echelle Spectrometer (HIRES) suggested an even lower mass of $30.5\pm1.7~\Mearth$~\citep{Piaulet2020}. This low density challenges the standard core-accretion model of planet formation. If runaway accretion brought WASP-107b to a gas-to-core mass ratio of $\sim 3$ but was stopped prematurely before growing to gas giant size, orbital dynamics and/or migration may have played a significant role in this system~\citep{Piaulet2020}. Alternatively WASP-107b's radius may be inflated from tidal heating, which would allow a lower gas-to-core ratio consistent with core accretion~\citep{Millholland2020}.

With a low density, large radius, and hot equilibrium temperature, WASP-107b's large atmospheric scale height makes it a prime target for atmospheric studies. Indeed analyses of transmission spectra obtained with the Hubble Space Telescope (HST)/WFC3 have detected water amongst a methane-depleted atmosphere~\citep{Kreidberg18}. WASP-107b was the first exoplanet to be observed transiting with excess absorption at 10830~\AA, an absorption line of a metastable state of neutral helium indicative of an escaping atmosphere~\citep{Oklopcic18}. These observations suggest that WASP-107b's atmosphere is photoevaporating at a rate of a few percent in mass per billion years \citep{Spake18, Allart2019, Kirk2020}.

The orbit of WASP-107b is suspected to be misaligned with the rotation axis of its host star. The angle between the star's rotation axis and the normal to the planet's orbital plane, called the stellar obliquity $\psi$ (or just obliquity), was previously constrained by observations of WASP-107b passing over starspots as it transited~\citep{DaiWinn17}. As starspots are regions of reduced intensity on the stellar photosphere that rotate with the star, this is seen as a bump of increased brightness in the transit light curve. By measuring the time between spot-crossing events across successive transits, combined with the absence of repeated spot crossings, \citet{DaiWinn17} were able to constrain the sky-projected obliquity, $\lambda$, of WASP-107b to $\lambda \in $[40--140]~deg. Intriguingly, long-baseline RV monitoring of the system with Keck/HIRES has revealed a distant ($P_c \sim 1100$~days) massive ($M\sin\iorbc = 115 \pm 13~\Mearth$) planetary companion, which may be responsible for this present day misaligned orbit through its gravitational influence on WASP-107b~\citep{Piaulet2020}.

The sky-projected obliquity can also be measured spectroscopically. The Rossiter--McLaughlin (RM) effect refers to the anomalous Doppler-shift caused by a transiting planet blocking the projected rotational velocities across the stellar disk~\citep{McLaughlin1924, Rossiter1924}. If the planet's orbit is aligned with the rotation of the star (prograde), its transit will cause an anomalous redshift followed by an anomalous blueshift. A anti-aligned (retrograde) orbit will cause the opposite to occur.

Following the first obliquity measurement by \citet{Queloz2000}, the field saw measurements of 10 exoplanet obliquities over the next 8 years that were all consistent with aligned, prograde orbits. After a few misaligned systems had been discovered \citep[e.g.,][]{Hebrard2008}, a pattern emerged with hot Jupiters on highly misaligned orbits around stars hotter than about $6250$~K~\citep{Winn2010hotStars}. This pattern elicited several hypotheses such as damping of inclination by the convective envelope of cooler stars~\citep{Winn2010hotStars} or magnetic realignment of orbits during the T Tauri phase~\citep{Spalding2015}. 

More recently a number of exoplanets have been found on misaligned orbits around cooler stars, such as the hot Jupiter WASP-8b~\citep{Queloz10, Bourrier2017}, as well as lower-mass hot Neptunes like
HAT-P-11b~\citep{Winn2010}, % might be most similar
% Kepler-56b~\citep{Huber2013}, % close to polar, but compact 2 inner planets. Is an evolved star, so really doesn't belong in this sample.
Kepler-63b~\citep{SanchisOjeda2013}, % -110deg and 30 Mearth
HAT-P-18b~\citep{Esposito2014}, % λ = 132 ± 15, Teff=4870 K
GJ 436b~\citep{Bourrier18},
and HD 3167 c~\citep{Dalal19}. % also compact multi
Strikingly, all of these exoplanets are on or near polar orbits. Some of these systems have recently had distant, giant companions detected \citep[e.g. HAT-P-11c;][]{Yee18}, hinting that these obliquities arise from multibody planet-planet dynamics.

In this paper we present a determination of the obliquity of WASP-107b from observations of the RM effect (Section~\ref{sec:obs}). These observations were acquired under the TESS--Keck Survey (TKS), a collaboration between scientists at the University of California, the California Institute of Technology, the University of Hawai`i, and NASA. TKS is organized through the California Planet Search with the goal of acquiring substantial RV follow-up observations of planetary systems discovered by TESS~\citep{Dalba2020}. TESS observed four transits of WASP-107b (TOI 1905) in Sector 10. An additional science goal of TKS is to measure the obliquities of interesting TESS systems. WASP-107b, which is already expected to have a significant obliquity~\citep{DaiWinn17}, is an excellent target for an RM measurement with HIRES.
 
In Section~\ref{sec:analysis} we confirm a misaligned orientation; in fact, we found a polar/retrograde orbit. This adds WASP-107b to the growing population of hot Neptunes in polar orbits around cool stars. We explored possible mechanisms that could be responsible for this misalignment in Section~\ref{sec:dynamics}. Lastly in Section~\ref{sec:conclusion} we summarized our findings and discussed the future work needed to better understand the obliquity distribution for small planets around cool stars.

% Our RVs
\begin{deluxetable}{lccc}
\tablecaption{Radial Velocities of WASP-107\label{tab:rvs}}
\tablehead{
  \colhead{Time} & 
  \colhead{RV}   & 
  \colhead{$\sigma_{\rm{RV}}$} & 
  \colhead{Exposure time} \\
  \colhead{(BJD$_{\text{TDB}}$)} &
  \colhead{(m s$^{-1}$)} & 
  \colhead{(m s$^{-1}$)} & 
  \colhead{(s)}
}
\startdata
2458905.90111  &  5.05 & 1.50 & 900 \\
2458905.91189  &  6.43 & 1.42 & 883 \\
2458905.92247  &  0.14 & 1.49 & 862 \\
2458905.93288  & -1.35 & 1.65 & 844 \\
2458905.94266  & -0.25 & 1.45 & 783 \\
2458905.95204  & -5.28 & 1.44 & 745 \\
2458905.96141  & -2.40 & 1.37 & 797 \\
2458905.97098  & -3.40 & 1.46 & 754 \\
2458905.98004  &  2.45 & 1.37 & 727 \\
2458905.98927  & -5.52 & 1.45 & 780 \\
2458905.99888  &  2.07 & 1.48 & 792 \\
2458906.00848  &  4.21 & 1.37 & 776 \\
2458906.01796  & -0.58 & 1.38 & 775 \\
2458906.02768  &  0.83 & 1.47 & 817 \\
2458906.03780  &  3.07 & 1.49 & 836 \\
2458906.04780  & -3.01 & 1.26 & 818 \\
2458906.05771  &  0.02 & 1.45 & 796 \\
2458906.06752  & -3.72 & 1.49 & 795 \\
2458906.07703  &  3.61 & 1.33 & 773 \\
2458906.08654  &  1.27 & 1.38 & 790 \\
2458906.09648  & -2.88 & 1.45 & 837 \\
2458906.10657  & -5.39 & 1.44 & 818 \\
\enddata
\tablecomments{A machine readable version is available.}
\end{deluxetable}

\section{Observations}\label{sec:obs}
% 02/26/2020 10:36:50j365522    EPIC228724232  y  10:43:02   745  (10:43:02) C2  60k  1.1"   2%
We observed the RM effect for WASP-107b during a transit on 2020 February 26 (UTC) with HIRES~\citep{Vogt1994} on the Keck I Telescope on Maunakea. Our HIRES observations covered the full transit duration ($\sim 2.7$~hr) with a $\sim 1$~hour baseline on either side. We used the ``C2'' decker ($14'' \times 0\farcs861$, $R = 45,000$) and integrated until the exposure meter reached 60,000 counts (signal-to-noise ratio (S/N) $\sim 100$ per reduced pixel, $\lesssim 15$ minutes) or readout after 15 minutes. The spectra were reduced using the standard procedures of the California Planet Search \citep{Howard2010}, with the iodine cell serving as the wavelength reference~\citep{Butler1996}. In total we obtained 22 RVs, 12 of which were in transit (Table~\ref{tab:rvs}).

% Table of adopted parameters
\begin{deluxetable}{lrrr}
\centering
\tablecaption{Adopted parameters of the WASP-107 System\label{tab:systemprops}}
\tablehead{
  \colhead{Parameter} & 
  \colhead{Value} & 
  \colhead{Unit} &
  \colhead{Source} 
}
\startdata
$P_b$           & $5.7214742$                &  days    & 1    \\
$t_c$           & $7584.329897 \pm 0.000032 $& JD\tablenotemark{a} & 1    \\
$b$             & $0.07 \pm 0.07$            &          & 1    \\ 
$\iorbb$        & $89.887^{+0.074}_{-0.097}$& degrees  & 1    \\
$R_p/R_\star$   & $0.14434 \pm 0.00018$    &            & 1    \\
$a / R_\star$   & $18.164 \pm 0.037$       &           & 1    \\
$e_b$           & $0.06 \pm 0.04$          &            & 2  \\
$\omega_b$      & $40^{+40}_{-60}$         & degrees    & 2  \\
$M_b$           & $30.5 \pm 1.7$           & M$_\oplus$ & 2  \\
\hline
$P_c$          & $1088^{+15}_{-16}$        & days        & 2 \\
$e_c$          & $0.28\pm 0.07$            &             & 2 \\
$\omega_c$     & $-120^{+30}_{-20}$        & degrees     & 2 \\
$M_c\sin\iorbc$& $0.36 \pm 0.04$   & M$_J$       & 2 \\
\hline
$T_{\text{eff}}$ & $4245\pm 70$              & K         & 2 \\
$M_\ast$         & $0.683^{+0.017}_{-0.016}$ & M$_\odot$ & 2 \\
$R_\ast$         & $0.67\pm0.02$             & R$_\odot$ & 2 \\
$u_1$            & $0.6666 \pm 0.0062$       &           & 1 \\
$u_2$            & $0.0150 \pm 0.0110$       &           & 1 \\
\enddata
\tablenotetext{a}{Days since JD 2,450,000. Sources: (1) \citet{DaiWinn17}; (2) \citet{Piaulet2020}.}
\end{deluxetable}

\begin{figure*}
    \centering
    \includegraphics[width=0.9\textwidth]{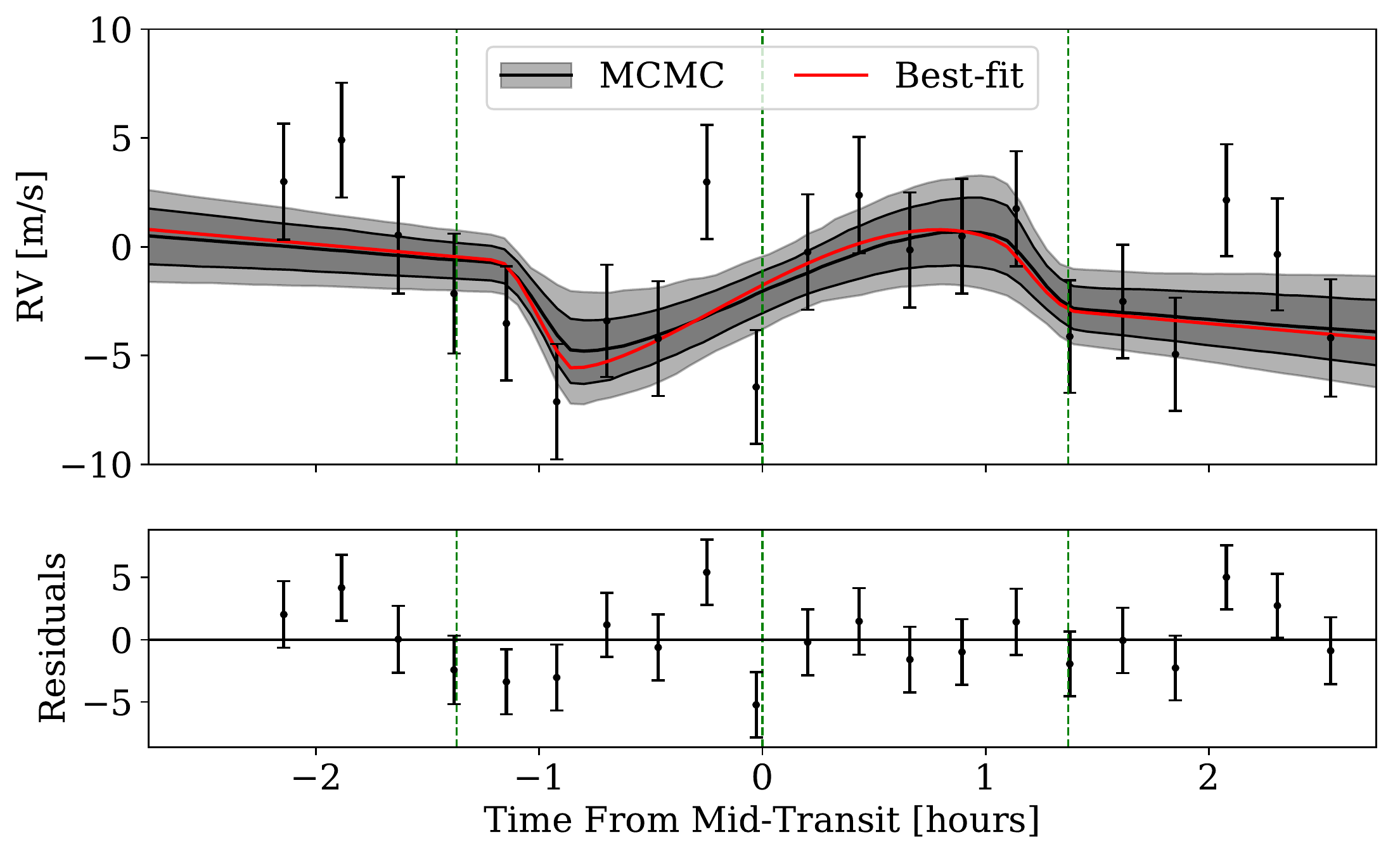}
    \caption{The RM effect for WASP-107b. The dark shaded bands show the 16th--84th (black) and 5th--95th (gray) percentiles from the posterior distribution of the modeled RV. The red best-fit line is the maximum a-posteriori (MAP) model. The three vertical dashed lines denote, in chronological order, the times of transit ingress, midpoint, and egress. The residuals show the data minus the best-fit model. Data points are drawn with the measurement errors and the best-fit jitter added in quadrature.}
    \label{fig:RM_fit}
\end{figure*}

Visually inspecting the observations~(Fig.~\ref{fig:RM_fit}) shows an anomalous blueshift following the transit ingress, followed by an anomalous redshift after the transit midpoint,\footnote{Propagating the uncertainty in $t_c$ in Table~\ref{tab:systemprops} the transit midpoint on the night of observation is uncertain to about 9~s.}, indicating a retrograde orbit. The asymmetry and low-amplitude of the signal constrain the orientation to a near-polar alignment, but whether the orbit is polar or anti-aligned is somewhat degenerate with the value of $\vsini$. The expected RM amplitude is $ \vsini (R_p/R_\star)^2 \sim 40$~m~s$^{-1}$, using previous estimates of $R_p/R_\star = 0.144$~\citep{DaiWinn17} and $\vsini \sim 2$~km~$s^{-1}$~\citep[e.g.,][]{Anderson2017}. The signal we detected with HIRES is only $\sim 5.5$~m~s$^{-1}$ in amplitude. \citet{DaiWinn17} found the transit impact parameter to be nearly zero, therefore the small RM amplitude suggests either a much lower $\vsini$ than was spectroscopically inferred (see Section~\ref{sec:stellar inclination}), a near-polar orbit, or both.

% RM analysis
\section{Analysis}\label{sec:analysis}

\subsection{Rossiter--McLaughlin Model}\label{sec:RM model}
We used a Gaussian likelihood for the RV time series $(\bm{t},\, \bm{v}_r)$ given the model parameters $\bm{\Theta}$, and included a RV jitter term ($\sigma_j$) to account for additional astrophysical or instrumental noise,
\begin{equation}\label{eq:likelihood}
p(\bm{v}_r,\,\bm{t} | \bm{\Theta}) = \prod_{i=1}^{N} \frac{1}{\sqrt{2\pi\sigma^2}} \exp\left[ -\frac{(v_{r,i} - f(t_i,\, \bm{\Theta}))^2}{2\sigma_i^2}\right],
\end{equation}
where $\sigma_i^2 = \sigma_\text{RV,i}^2 + \sigma_j^2$. The model $f(t_i,\, \bm{\Theta})$ is given by
\begin{equation}\label{eq:rm-model}
f(t_i,\, \bm{\Theta}) = \text{RM}(t_i,\,\bm{\theta}) + \gamma + \dot{\gamma}(t_i - t_0),
\end{equation}
where $\bm{\Theta} = (\bm{\theta},\, \gamma,\,\dot{\gamma})$ is the RM model parameters ($\bm{\theta}$) as well as an offset ($\gamma$) and slope ($\dot{\gamma}$) term which we added to approximate the reflex motion of the star and model any other systematic shift in RV throughout the transit (e.g., from noncrossed spots). The reference time $t_0$ is the time of the first observation (BJD).

$\text{RM}(t_i,\,\bm{\theta})$ is the RM model described in \citet{Hirano11}. We assumed zero stellar differential rotation and adopted the transit parameters determined by \cite{DaiWinn17}, which came from a detailed analysis of K2 short-cadence photometry. 
We performed a simultaneous fit to the photometric and spectroscopic transit data using the same photometric data from K2 as in \citet{DaiWinn17} to check for consistency. We obtained identical results for the transit parameters as they did, hence we opted to simply adopt their values, including their quadratic limb-darkening model. These transit parameters are all listed in Table~\ref{tab:systemprops}. Our best-fit RV jitter is $\sigma_j = 2.61_{-0.51}^{+0.64}$~m~s$^{-1}$, smaller than the jitter from the Keplerian fit to the full RV sample of $3.9_{-0.4}^{+0.5}$~m~s$^{-1}$~\citep{Piaulet2020}. This is expected as the RM sequence covers a much shorter time baseline as compared to the full RV baseline, and as a result is only contaminated by short-term stellar noise sources such as granulation and convection.

The free parameters in the RM model are the sky-projected obliquity ($\lambda$), stellar inclination angle ($\istar$), and projected rotational velocity ($\vsini$). To first order, the impact parameter $b$ and sky-projected obliquity $\lambda$ determine the shape of the RM signal, while $\vsini$ and $R_p/R_\star$ set the amplitude. We adopted the parameterization ($\sqrt{\vsini}\cos\lambda$, $\sqrt{\vsini}\sin\lambda$) to improve the sampling efficiency and convergence of the Markov Chain Monte Carlo (MCMC). A higher order effect that becomes important when the RM amplitude is small is the convective blueshift, which we denote $v_{cb}$ (see Section~\ref{sec:CB} for more details). There are thus seven free parameters in our model: $\sqrt{\vsini}\cos\lambda$, $\sqrt{\vsini}\sin\lambda$, $\cos\istar$, $\log(|v_{cb}|)$, $\gamma$, $\dot{\gamma}$, and $\sigma_j$. We placed a uniform hard-bounded prior on $\vsini \in [0,\,5]$~km~s$^{-1}$ and on $\cos\istar \in [0,\,1]$, and used a Jeffrey's prior for $\sigma_j$. All other parameters were assigned uniform priors.

\subsection{Micro/Macroturbulence Parameters}\label{sec:turbulence}
The shape of the RM curve is also affected by processes on the surface of the star that broaden spectral lines, which affect the inferred RVs. In the \citet{Hirano11} model, these processes are parameterized by $\gamma_{\text{lw}}$, the intrinsic line width, $\zeta$, the line width due to macroturbulence, given by the \citet{ValentiFischer2005} scaling relation
\begin{equation}\label{eq:zeta}
\zeta = \left(3.98 + \frac{T_\text{eff} - 5770~\text{K}}{650~\text{K}}\right)~\text{km s}^{-1},
\end{equation}
and $\beta$, given by
\begin{equation}\label{eq:beta}
\beta = \sqrt{\frac{2 k_B T_{\text{eff}}}{\mu} + \xi^2 + \beta_{\text{IP}} },
\end{equation}
where $\xi$ is the dispersion due to microturbulence and $\beta_{\text{IP}}$ is the Gaussian dispersion due to the instrument profile, which we set to the HIRES line-spread function (LSF) (2.2~km~s$^{-1}$). 
We tested having $\gamma_{\text{lw}},\, \xi$, and $\zeta$ as free parameters in the model (with uniform priors) but only recovered the prior distributions for these parameters. Moreover we saw no change in the resulting posterior distribution for $\lambda$ or $\vsini$. Because of this, we opted to instead adopt fixed nominal values of $\xi = 0.7$~km~s$^{-1}$, $\gamma_{\text{lw}} = 1$~km~s$^{-1}$, and $\zeta = 1.63$~km~s$^{-1}$ (from Eq.~\ref{eq:zeta} using $T_{\text{eff}}$ from Table~\ref{tab:systemprops}).

\subsection{Convective blueshift}\label{sec:CB}

Convection in the stellar photosphere, caused by hotter bubbles of gas rising to the stellar surface and cooler gas sinking, results in a net blueshift across the stellar disk. This is because the rising (blueshifted) gas is hotter, and therefore brighter, than the cooler sinking (redshifted) gas. Since this net-blueshifted signal is directed at an angle normal to the stellar surface, the radial component seen by the observer is different in amplitude near the limb of the star compared to the center of the stellar disk, according to the stellar limb-darkening profile. Thus the magnitude of the convective blueshift blocked by the planet varies over the duration of the transit. The amplitude of this effect is $\sim 2$~m~s$^{-1}$, which is significant given the small amplitude of the RM signal we observe for WASP-107b ($\sim 5.5$~m~s$^{-1}$). 

For this reason we included the prescription of \cite{ShporerBrown2011} in the RM model, which is parameterized by the magnitude of the convective blueshift integrated over the stellar disk ($v_{cb}$). This quantity is negative by convention. Since the possible value of $v_{cb}$ could cover several orders of magnitude, we fit for $\log(|v_{cb}|)$ and set a uniform prior between -1 and 3. While we found that including $v_{cb}$ has no effect on the recovered $\lambda$ and $\vsini$ posteriors, we are able to rule out $|v_{cb}| > \Vcbnn$~m~s$^{-1}$ at 99\% confidence, and $>\Vcbnf$~m~s$^{-1}$ at 95\% confidence.

\begin{figure}
    \centering
    \includegraphics[width=0.48\textwidth]{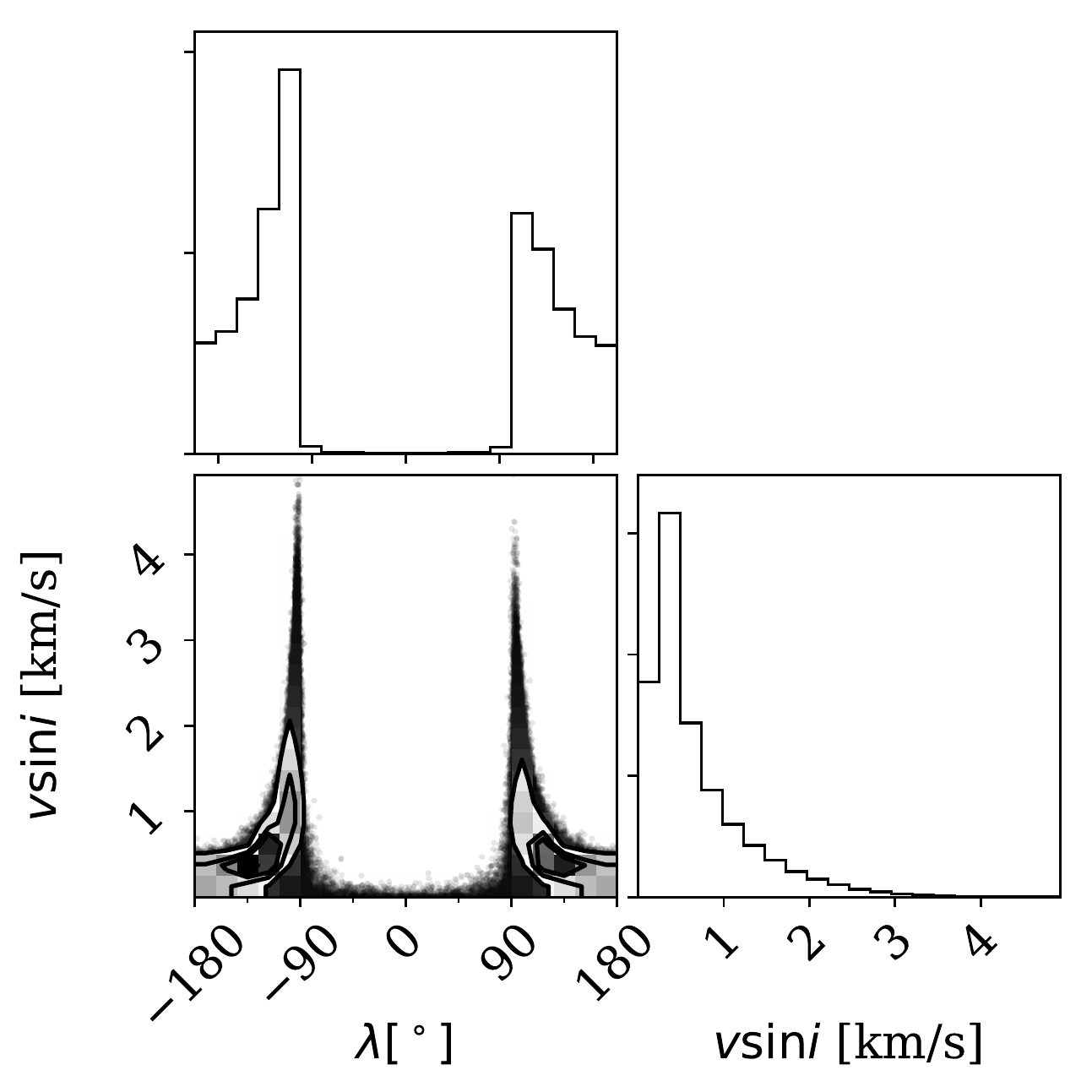}
    \caption{Posterior distribution for $\lambda$ and $\vsini$. Although a more anti-aligned configuration is consistent with the data if $\vsini$ is small, the most likely orientations are close to polar. A prograde orbit ($|\lambda|<90^\circ$) is strongly ruled out.}
    \label{fig:cornerplot}
\end{figure}
\begin{figure}
    \centering
    \includegraphics[width=0.48\textwidth]{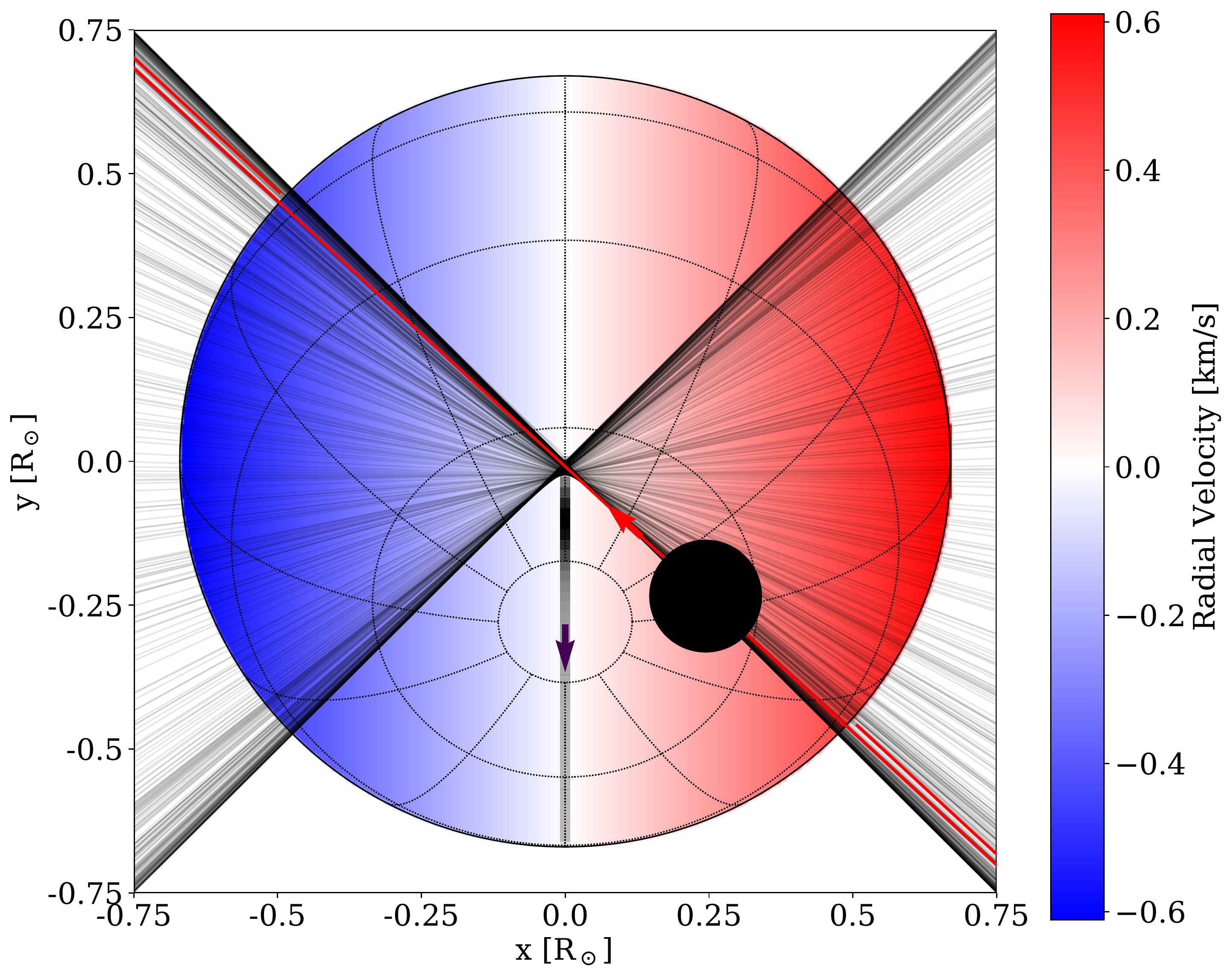}
    \caption{Sky-projected orbital configuration of WASP-107b's orbit relative to the stellar rotation axis. The black lines correspond to posterior draws while the red line is the MAP orbit from Fig.~\ref{fig:RM_fit}. The direction of WASP-107b's orbit is denoted by the red arrow. The stellar rotation axis (black arrow) and lines of stellar latitude and longitude are drawn for an inclination of $\istar = 25^\circ$. The posterior for $\istar$ is illustrated by the shaded gray strip with a transparency proportional to the probability.}
    \label{fig:orbit_on_sky_posterior}
\end{figure}

\subsection{Evidence for a Retrograde/Polar Orbit}

% Table of MCMC/MAP fit results
\begin{deluxetable}{lrrr}
\centering
\tablecaption{WASP-107b Rossiter--McLaughlin Parameters\label{tab:mcmcresults}}
\tablehead{
  \colhead{Parameter} & 
  \colhead{MCMC CI} & 
  \colhead{MAP value} &
  \colhead{Unit}
}
\startdata
\multicolumn{4}{l}{Model~Parameters} \\
$\sqrt{\vsini}\cos\lambda$ & $-0.309_{-0.154}^{+0.150}$     & -0.30  & \tablenotemark{a}  \\
$\sqrt{\vsini}\sin\lambda$ & $-0.126_{-0.771}^{+0.808}$     & -0.72  & \tablenotemark{a}  \\
$\cos i_s$                 & $-0.003_{-0.681}^{+0.682}$     & -0.56  &             \\
$\gamma$                   & $0.80_{-1.38}^{+1.36}$      & 0.97   &  m s$^{-1}$ \\
$\dot{\gamma}$             & $-20.83_{-10.94}^{+11.05}$  & -21.85 &  m s$^{-2}$ \\
$\sigma_{\text{jit}}$      & $2.61_{-0.51}^{+0.64}$      & 2.20   &  m s$^{-1}$ \\
$\log(|v_{cb}|)$           & $0.89_{-1.27}^{+1.18}$      & 2.17   & \tablenotemark{a}  \\
\hline
\multicolumn{4}{l}{Derived~Parameters} \\
% $\lambda$  & $196.419_{-85.060}^{+60.316}$  & -112.63  &  degrees     \\
$|\lambda|$& $118.1_{-19.1}^{+37.8}$        & 112.63   &  degrees     \\
$\vsini$   & $0.45_{-0.23}^{+0.72}$         & 0.61     &  km s$^{-1}$ \\
$v_{cb}$   & $-7.74_{-109.71}^{+7.33}$      & -149.41  &   m s$^{-1}$ \\
$i_\star$  & $28.17_{-20.04}^{+40.38}$      & 7.06     &  degrees     \\
$|\psi|$   & $109.81_{-13.64}^{+28.17}$     & 92.60    &  degrees     \\
\enddata
\tablenotetext{a}{
$\vsini$ is in km~s$^{-1}$ and $v_{cb}$ is in m~s$^{-1}$.
}
\end{deluxetable}

We first found the maximum a posteriori (MAP) solution by minimizing the negative log-posterior using Powell’s method \citep{Powell1964} as implemented in \texttt{scipy.optimize.minimize}~\citep{scipy}. The MAP solution was then used to initialize an MCMC. We ran 8 parallel ensembles each consisting of 32 walkers for 10,000 steps using the python package \texttt{emcee} \citep{emcee}. We checked for convergence by requiring that both the Gelman--Rubin statistic~\citep[G--R;][]{Gelman2003} was $< 1.001$ across the ensembles~\citep{Ford2006} and the autocorrelation time was $< 50$ times the length of the chains~\citep{emcee}.

% The out-of-transit RV trend is consistent with the Keplerian prediction given the orbital parameters of WASP-107b, but is shifted low by about 1--2$\sigma$. This may be due to spots or other features on the surface of the star affecting the net stellar RV. Indeed spots have been previously observed during photometric transits of WASP-107b~\citep{DaiWinn17}.

The MAP values and central 68\% confidence intervals (CI) computed from the MCMC chains are tabulated in Table~\ref{tab:mcmcresults}, and the full posteriors for $\lambda$ and $\vsini$ are shown in Fig.~\ref{fig:cornerplot}. A prograde ($|\lambda| < 90^\circ$) orbit is ruled out at $>99\%$ confidence. An anti-aligned ($135^\circ < \lambda < 225^\circ$) orbit is allowed if $\vsini$ is small ($0.26 \pm 0.10$~km~s$^{-1}$), although a more polar aligned (but still retrograde) orbit with $90^\circ < |\lambda| < 135^\circ$ is more likely (if $\vsini \in [0.22,\,2.09]$~km~s$^{-1}$, 90\% CI). The true obliquity $\psi$ will always be closer to a polar orientation than $\lambda$, since $\lambda$ represents the minimum obliquity in the case where the star is viewed edge-on ($\istar = 90^\circ$). While an equatorial orbit that transits requires $\istar \sim 90^\circ$, a polar orbit may be seen to transit for any stellar inclination.

To confirm that the signal we detected was not driven by correlated noise structures in the data, we performed a test using the cyclical residual permutation technique. We first calculated the residuals from the MAP fit to the original RV time series. We then shifted these residuals forward in time by one data point, wrapping at the boundaries, and added these new residuals back to the MAP model. This new ``fake'' dataset was then fit again and the process was repeated $N$ times where $N = 22$ is the number of data points in our RV time series. This technique preserves the red noise component, and permuting multiple times generates datasets that have the same temporal correlation but different realizations of the data. If we assume that the signal we detected is caused by a correlated noise structure, then we would expect to see the detected signal vanish or otherwise become significantly weaker across each permutation as that noise structure becomes asynchronous with the transit ephemeris. We found that the signal is robustly detected at all permutations, with and without including the convective blueshift (fixed to the original MAP value). The MAP estimate for $\lambda$ tended to be closer to polar across the permutations compared as to the original fit, which is consistent with the posterior distribution estimated from the MCMC, but did not vary significantly. While this method is not appropriate for estimating parameter uncertainties~\citep{Cubillos2017}, we conclude that our results are not qualitatively affected by correlated noise in our RV time series.

Spot-crossing events can also affect the RM curve since the planet would block a different amount of red/blueshifted light. Out of the nine transits observed by \citet{DaiWinn17}, a single spot-crossing event was seen in only three of the transits. Hence there is roughly a one in three chance that the transit we observed contained a spot-crossing event. As we did not obtain simultaneous high-cadence photometry, we do not know if or when such an event occurred. Judging from the durations ($\sim 30$ min) of the spot crossings observed by \citet{DaiWinn17}, this would only affect one or maybe two of our 15-minute exposures. While we don’t see any significant outliers in our dataset, these spots were only $\sim 10\%$ changes on a  $\sim 2\%$ transit depth, amounting to an overall spot depth of $\sim 0.2\%$. Given our estimate of $\vsini \sim 0.5$~km~s$^{-1}$ this suggests a spot-crossing event would produce a $\sim 1$~m~s$^{-1}$ RV anomaly, small compared to our measurement uncertainties ($\sim 1.5$~m~s$^{-1}$) and the estimated stellar jitter ($\sim 2.6$~m~s$^{-1}$). In other words, there is a roughly 33\% chance that a spot-crossing event introduced an additional $0.5\sigma$ error on a single data point. If there were multiple spot-crossing events this anomaly would vary across the transit similar to other stellar-activity processes. In practice this introduces a correlated noise structure in the RV time series which our cyclical residual permutation test demonstrated is not significantly influencing our measurement of the obliquity or other model parameters. From this semi-analytic analysis we conclude that spot crossings are not a leading source of uncertainty in our model.

\subsection{Constraints on the Stellar Inclination}\label{sec:stellar inclination}

\begin{figure}
    \centering
    \includegraphics[width=0.495\textwidth]{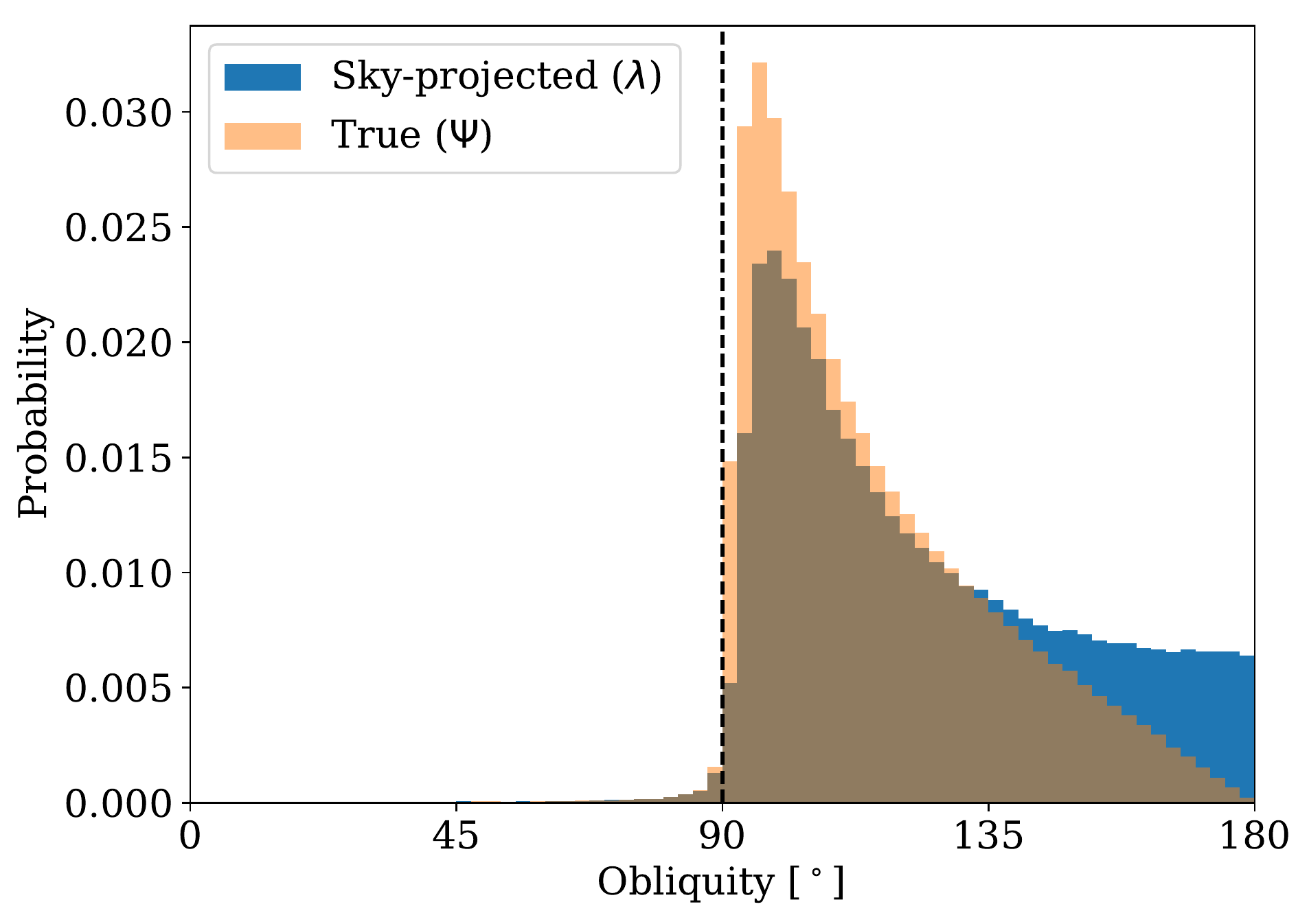}
    \caption{Obliquity of WASP-107b. The true obliquity $\psi$ is calculated using the constraints on the stellar inclination as inferred from the $\vsini$ posterior~(Section~\ref{sec:stellar inclination}).}
    \label{fig:true obliquity}
\end{figure}

Given a constraint on $\vsini$ and $v$, we can constrain the stellar inclination $\istar$. Previous studies have found a range of estimates for the $\vsini$ of WASP-107. \cite{Anderson2017} found a value of $2.5 \pm 0.8$~km~s$^{-1}$, whereas John Brewer (private communication) obtained a value of $1.5 \pm 0.5$~km~s$^{-1}$ using the automated spectral synthesis modeling procedure described in~\citet{Brewer2016}. We note that the Specmatch-Emp~\citep{smemp} result for our HIRES spectrum only yields an upper bound for $\vsini$ of $< 2$~km~s$^{-1}$, as this technique is limited by the HIRES PSF. All three of these methods derive $\vsini$ by modeling the amount of line broadening present in the stellar spectrum, which in part comes from the stellar rotation. However these estimates may be biased from other sources of broadening which are not as well constrained in these models. Our RM analysis on the other hand incorporates a direct measurement of $\vsini$ by observing how much of the projected stellar rotational velocity is blocked by the transiting planet's shadow. Our RM analysis found $\vsini = 0.45_{-0.23}^{+0.72}$~km~s$^{-1}$, lower than the spectroscopic estimates. We adopted this posterior for $\vsini$ to keep internal consistency.

The rotation period of WASP-107 has been estimated to be $17 \pm 1$~days from photometric modulations due to starspots rotating in and out of view~\citep{Anderson2017, DaiWinn17, Mocnik17}. We combined this rotation period with the stellar radius of $0.67 \pm 0.02~R_\odot$ inferred from the HIRES spectrum~\citep{Piaulet2020} using Specmatch-Emp~\citep{smemp} to constrain the tangential rotational velocity $v = 2\pi R_\star / P_\text{rot}$. We then used the statistically correct procedure described by \cite{MasudaWinn2020} and performed an MCMC sampling of $v$ and $\cos\istar$, using uniform priors for each, and using the posterior distribution for $\vsini$ obtained in the RM analysis as a constraint. Sampling both variables simultaneously correctly incorporates the nonindependence of $v$ and $\cos\istar$, since $v \leq \vsini$. We found that $\istar = 25.8_{-15.4}^{+22.5}$ degrees (MAP value $7.1^\circ$), implying a viewing geometry of close to pole-on for the star. Thus any transiting configuration will necessarily imply a near-polar orbit, even for orbital solutions with $\lambda$ near $180^\circ$ (see Fig.~\ref{fig:orbit_on_sky_posterior}). It is worth mentioning that one of the three spot-crossing events observed by \citet{DaiWinn17} occurred near the transit midpoint. This small stellar inclination implies that this spot must be at a relatively high latitude ($90^\circ - \istar$) compared to that of our Sun, which has nearly all of its sunspots contained within $\pm30^\circ$ latitude.

Knowledge of the stellar inclination $\istar$, the orbital inclination $\iorb$, and the sky-projected obliquity $\lambda$ allows one to compute the true obliquity $\psi$, as these four angles are related by
\begin{equation}\label{eq:true obliquity} % https://link.springer.com/chapter/10.1007%2F978-981-10-8453-9_2
    \cos\psi = \cos\iorb \cos\istar + \sin\iorb \sin\istar \cos\lambda.
\end{equation}
The resulting posterior distribution for the true obliquity $\psi$ is shown in Fig.~\ref{fig:true obliquity}. As expected, the true orbit is constrained to a more polar orientation than is implied by the wide posteriors on $\lambda$, due to the nearly pole-on viewing geometry of the star itself.

% Dynamical history/formation
\section{Dynamical History}\label{sec:dynamics}

%Blurb from Yee et al. 2018: Many explanations have been proposed for such misalignments, including Kozai– Lidov cycles (e.g., Fabrycky & Tremaine 2007), planet–planet scattering (e.g., Nagasawa et al. 2008), primordial tilting of the protoplanetary disk (e.g., Batygin 2012), or angular momentum transport by internal gravity waves (e.g., Rogers et al. 2012).

% Want casual reader to get an idea of it, and sophisticated reader to be able to reproduce it. Find a balance between equations/derivation and descriptive wording w/ references to Yee et al. and Xuan et al.
How did WASP-107b end up in a slightly retrograde, nearly polar orbit? To explore this question, we examined the orbital dynamics of the WASP-107 system considering the new discovery of a distant, giant companion WASP-107c~\citep{Piaulet2020}. As in \cite{Mardling2010}, \cite{Yee18}, and \cite{Xuan2020}, we can understand the evolution of the WASP-107 system by examining the secular three-body Hamiltonian. Assuming the inner planet is a test particle (i.e., $M_b\sqrt{a_b} \ll M_c\sqrt{a_c}$), and since $a_b/a_c \ll 1$, we can approximate the Hamiltonian by expanding to quadrupole order in semimajor axis ratio
\begin{eqnarray}\label{eq:Hamiltonian}
    \mathcal{H} &=& \frac{1}{16}n_b\frac{M_c}{M_\star}\left(\frac{a_b}{a_c\sqrt{1-e_c^2}}\right)^3 \left[\frac{(5-3G_b^2)(3H_b^2-G_b^2)}{G_b^2}\right. \nonumber\\
    &&\left.+ \frac{15(1-G_b^2)(G_b^2-H_b^2)\cos(2g_b)}{G_b^2} \right] + \frac{GM_\star}{a_b c^2}\frac{3n_b}{G_b},
\end{eqnarray}
where the last term is the addition from general relativity (GR) and $n_b = 2\pi/P_b$. The quantities $G$ and $H$ are the canonical Delaunay variables
\begin{align}\label{eq:kozai constant}
    G_b &= \sqrt{1 - e_b^2} & \hspace{-50pt} \leftrightarrow \; g_b = \omega_b, \\
    H_b &= G\cos i_b        & \hspace{-50pt} \leftrightarrow \; h_b = \Omega_b, \nonumber
\end{align}
where the double-arrow ($\leftrightarrow$) symbolizes conjugate variables, $\omega_b$ is the argument of perihelion of the inner planet, $\Omega_b$ is the longitude of ascending node of the inner planet, and $i_b$ is the inclination of the inner planet with respect to the invariant plane. The invariant plane is the plane normal to the total angular momentum bmtor, which to good approximation is simply the orbital plane of the outer planet (since angular momentum is $\propto M a^{1/2}$). With this approximation, $i_b$ is the relative inclination between the two planets.

\begin{figure*}[t]
    \centering
    \includegraphics[width=\textwidth]{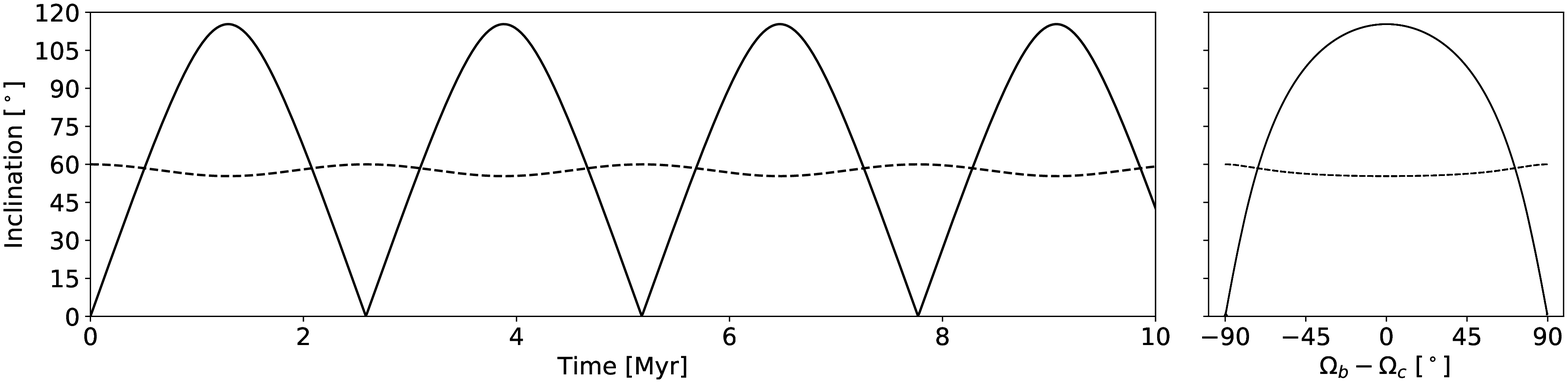}
    % \caption{Evolution of $e_b$ and $\psi_b$ (solid lines) throughout the the $N$-body simulation using the system parameters given in Table~\ref{tab:systemprops}. The outer planet has $M_c = M\sin\iorbc$ and was initialized with an obliquity of $\psi_c = 50^\circ$ (dashed line). Any small initial eccentricity is quickly damped by tidal forces ($\tau_\text{circ} = e/\dot{e} \sim 3$~Myr), while the obliquity of planet b oscillates between $\psi_c \pm \psi_c$ every $\sim 2$~Myr due to nodal precession. If $\sin\iorbc < 1$ then the larger $M_c$ simply produces a shorter nodal precession timescale.}
    \caption{Evolution of WASP-107b's true obliquity ($\psi_b$, solid line) throughout the the $N$-body simulation using the system parameters given in Table~\ref{tab:systemprops}. The outer planet has $M_c = M\sin\iorbc$ and was initialized with an obliquity of $\psi_c = 60^\circ$ (dashed line). The obliquity of planet b oscillates between $\psi_c \pm \psi_c$ every $\sim 2.5$~Myr due to nodal precession. If $\sin\iorbc < 1$ then the larger $M_c$ simply produces a shorter nodal precession timescale. The right panel shows the evolution of the inclinations with the difference in the longitudes of ascending node.}
    \label{fig:nbody}
\end{figure*}

\subsection{Kozai--Lidov oscillations}\label{sec:kozai}

Since the Hamiltonian $\mathcal{H}$ does not depend on $h_b$, the quantity $H_b = \sqrt{1 - e_b^2}\cos i_b$ is conserved. This leads to a periodic exchange of $e_b$ and $i_b$, so long as the outer planet has an inclination greater than a critical value of $\sim 39\fdg2$~\citep{Kozai1962, Lidov1962}. These Kozai--Lidov cycles also require a slowly changing argument of perihelion, which may precess due to GR as is famously seen in the orbit of Mercury. This precession can suppress Kozai--Lidov cycles if fast enough, as is the case for HAT-P-11 and $\pi$~Men~\citep{Xuan2020, Yee18}. The precession rate from GR is given by
\begin{equation}\label{eq:GR precession}
    \dot{\omega}_{GR} = \frac{GM_\star}{a_b c^2}\frac{3n_b}{G_b^2},
\end{equation}
which has an associated timescale of $\tau_{GR} = 2\pi/\dot{\omega} \approx 42,500$ years for WASP-107b. The Kozai timescale \citep{Kiseleva1998} is
\begin{equation}\label{eq:kozai timescale}
    \tau_{\text{Kozai}} = \frac{2P_c^2}{3\pi P_b^2}\frac{M_\star}{M_c}(1 - e_c^2)^{3/2} \approx 210,000~\text{yr},
\end{equation}
five times longer. The condition for Kozai--Lidov cycles to be suppressed by relativistic precession is $\tau_\text{Kozai} \dot{\omega}_\text{GR} > 3$~\citep{Fabrycky2007}, which the MAP minimum mass and orbital parameters WASP-107c satisfy. This is nicely visualized in Figure~6 of Piaulet et al. (submitted), which shows the full posterior distributions of $\tau_\text{Kozai}$ and $\tau_\text{GR}$. While the true mass of WASP-107c is likely to be larger than the derived $M\sin\iorbc$, it would need to be $\sim 10$ times larger for Kozai--Lidov oscillations to occur. This would imply a near face-on orbit of at most $\iorbc < 5\fdg5$. Such a face-on orbit is unlikely but is still plausible if it is aligned with the rotation axis of the star, given our constraints on the stellar inclination angle in Section~\ref{sec:stellar inclination}.

\subsection{Nodal precession}\label{sec:nodal precession}

An alternative explanation for the high obliquity of WASP-107b is nodal precession, as was proposed for HAT-P-11b~\citep{Yee18} and for $\pi$~Men~c~\citep{Xuan2020}. In this scenario the outer planet must have an obliquity greater than half that of the inner planet, which in this case would require $\psi_\text{c} \sim 55^\circ$. Then the longitude of ascending node $\Omega_b$ evolves in a secular manner according to~\citet{Yee18},
\begin{equation}\label{eq:nodal precession rate}
    \frac{d\Omega_b}{dt} = \frac{\partial \mathcal{H}}{\partial H_b} = \frac{n_b}{8}\frac{M_c}{M_\star} \left(\frac{a_b}{a_c\sqrt{1-e_c^2}}\right)^3\left(\frac{15 - 9G_b^2}{G_b^2}\right)H_b.
\end{equation}
 The associated timescale $\tau_{\Omega_b} = 2\pi/\dot{\Omega}_b$ is only about 2~Myr, much shorter than the age of the system. \citet{Yee18} pointed out that such a precession will cause the relative inclination of the two planets to oscillate between $\approx \psi_c \pm \psi_c$. Thus at certain times the observer may see a highly misaligned orbit ($\psi_b \sim 2\psi_c$) for the inner planet, while at other times the observer may see an aligned orbit ($\psi_b = 0$).

We examined this effect by running a 3D $N$-body simulation in \texttt{REBOUND} \citep{rebound}. We initialized planet c with an obliquity of 60$^\circ$ (which sets the maximum obliquity planet b can obtain, ${\sim}2\psi_c = 120^\circ$) and planet b with an obliquity of 0$^\circ$ (aligned, prograde orbit). We included the effects of GR and tides using the \texttt{gr} and \texttt{modify\_orbits\_forces} features of \texttt{REBOUNDx}~\citep{Kostov2016, reboundx} and used the the WHFast integrator \citep{WHFast} to evolve the system forward in time for 10~Myr.

\begin{figure*}
    \centering
    \includegraphics[width=0.8\textwidth]{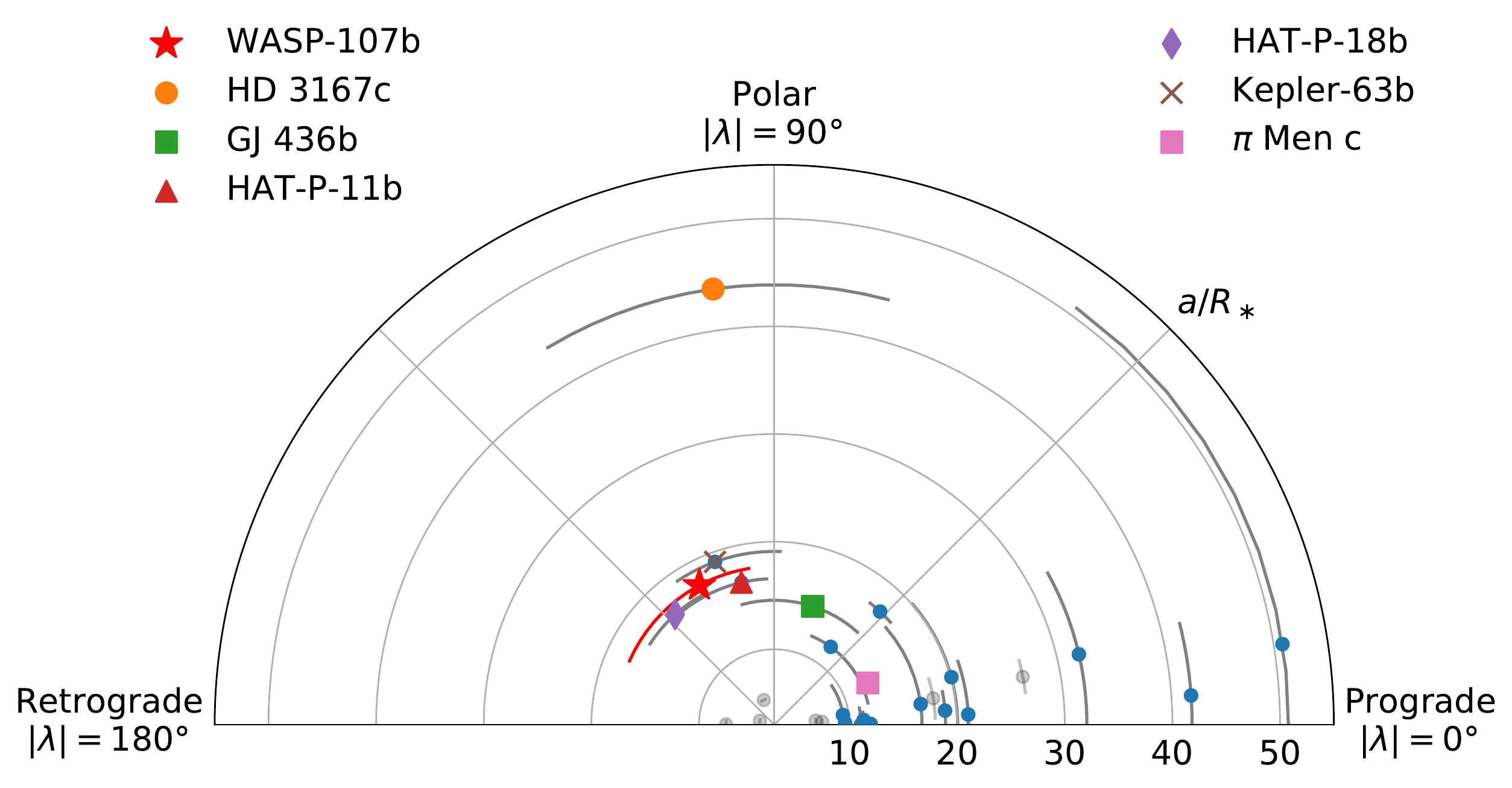}
    \includegraphics[width=0.8\textwidth]{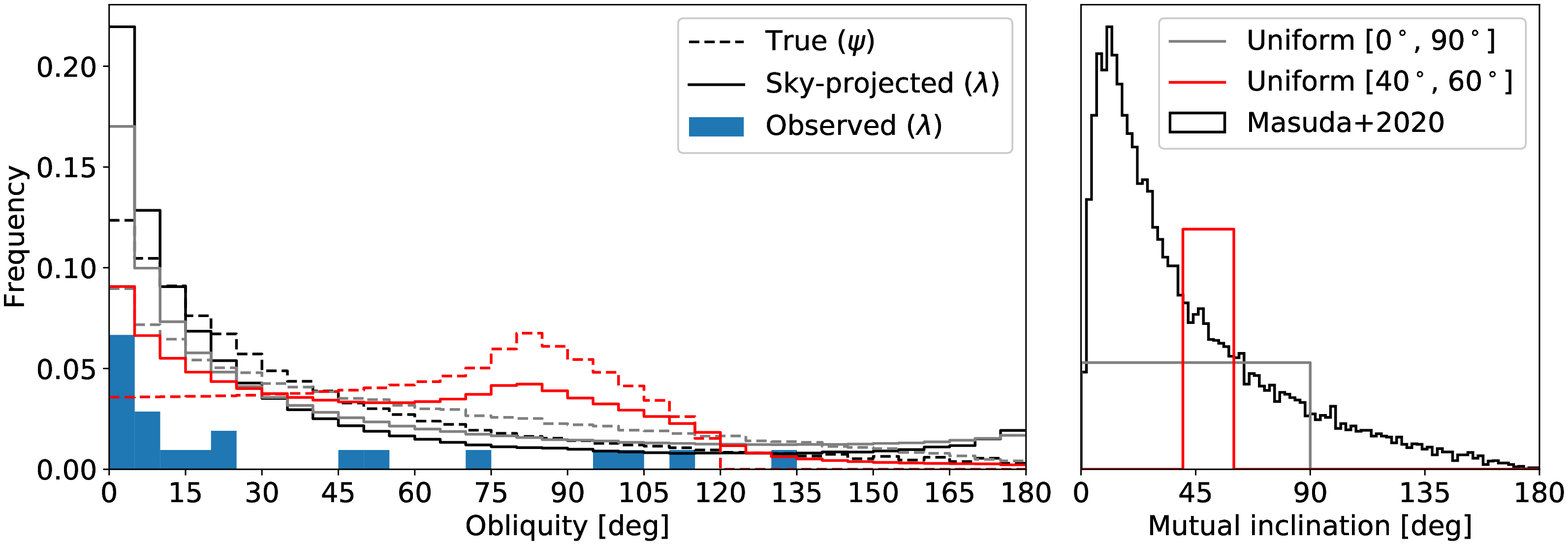}
    \caption{Top: polar plot showing the absolute sky-projected obliquity as the azimuthal coordinate and normalized orbital distance as the radial coordinate, for ${<}100~\Mearth$ planets around stars with $T_\text{eff} < 6250$~K (similar mass planets around hotter stars are shown as faded gray points). The red point is WASP-107b. Other noteworthy systems are shown with various colors and markers (see Section~\ref{sec:intro} for references). Data compiled from TEPCat as of 2020 October~\citep{Southworth2011}. Only WASP-107, HAT-P-11, and $\pi$~Men have distant giant companions detected. Kepler-56~\citep{Huber2013} is another similar system but is not included in this plot as it is an evolved massive star.
    Bottom: the fraction of a nodal precession cycle spent in a given obliquity bin (left). The true obliquity $\psi$ is assumed to vary as $\cos[(\pi/2)\psi(t)/\psi_\text{max}] = \sin^2(\pi t/\tau)$, where $t \in [0, \tau=1]$. This recreates the shape of the oscillating inclination in Fig.~\ref{fig:nbody}. The amplitude $\psi_\text{max}$ is twice the outer planet's inclination which is plotted for three different distributions (shown on the right): uniform between $[0^\circ,\, 90^\circ]$ (gray), uniform between $[40^\circ,\, 60^\circ]$ (red), and using the von-Mises Fisher distribution from \citet{Masuda2020} calculated in a hierarchical manner incorporating their posterior distribution for the shape parameter $\sigma$ for all. In all three cases the true obliquity is shown as a dashed histogram. The sky-projected obliquity is computed given a transiting geometry ($\iorbb = 90^\circ$) and is marginalized over stellar inclination angle (solid histogram). $M_p < 100~\Mearth$ planets with observed sky-projected obliquities are shown as a filled histogram for comparison. Note that while the gray and black predictions are relatively similar, an excess of polar orbits can be observed if the mutual inclination distribution is clustered around $\sim 40$--$60^\circ$.}
    \label{fig:obliquity demographics}
\end{figure*}

Fig.~\ref{fig:nbody} shows that over these 10~Myr $\psi_b$ oscillates in the range $0^\circ\text{--}120^\circ$ due to the precession of $\Omega_b$. Thus nodal precession can easily produce high relative inclinations, despite Kozai--Lidov oscillations being suppressed by GR. A configuration like what is observed today in which the inner planet is misaligned on a polar, yet slightly retrograde orbit is attainable at times during this cycle where the mutual inclination is at or near its maximum. The obliquity is $\gtrsim 80\%$ the amplitude from nodal precession (${\sim}2\psi_c$) approximately one-third of the time (bottom panel in Fig.~\ref{fig:obliquity demographics}). Therefore, even though the observed obliquity depends on when during the nodal precession cycle the system is observed, there is a decent chance of observing $\psi_b$ near its maximum. 

In the simulation we ran, WASP-107b is only seen by an observer to be in a transiting geometry about 2.8\% of the time. \citet{Xuan2020} did a more detailed calculating accounting for the measured mutual inclination and found that the dynamical transit probability for $\pi$~Men c and HAT-P-11b is of order 10-20\%. However, as \citet{Xuan2020} point out, this does not affect the population-level transit likelihood since the overall orientations of extrasolar systems can still be treated as isotropic. It merely suggests that a system with a transiting distant giant planet may be harboring a nodally precessing inner planet that just currently happens to be nontransiting.

Both Kozai--Lidov and nodal precession require a large mutual inclination in order for the inner planet to reach polar orientations. The origin of this large mutual inclination may be hidden in the planet's formation history, or perhaps was caused by a planet-planet scattering event with an additional companion that was ejected from the system. This could also explain the moderately eccentric orbit of WASP-107c~\citep{Piaulet2020}. Indeed a significant mutual inclination is observed for the inner and outer planets of the HAT-P-11 and $\pi$~Men systems \citep{Xuan2020}, although the inner planet in $\pi$~Men is only slightly misaligned with $\lambda = 24 \pm 4.1$~degrees~\citep{Hodzic2020}, while HAT-P-11b has $\lambda = 103_{-10}^{+26}$~degrees~\citep{Winn2010}.

As more close-in Neptunes with distant giant companions are discovered, the distribution of observed obliquities for the inner planet will help determine if we are indeed simply seeing many systems undergoing nodal precession but at different times during the precession cycle. If so, we might observe a sky-projected obliquity distribution that resembles the bottom panel of Fig.~\ref{fig:obliquity demographics}. However, we may instead be observing two classes of close-in Neptunes: ones aligned with their host stars and ones in polar or near-polar orbits (see the top panel of Fig.~\ref{fig:obliquity demographics}). This suggests an alternative mechanism that favors either polar orbits or aligned orbits depending on the system architecture.

\subsection{Disk dispersal-driven tilting}\label{sec:dddobliquity}

Recently, \citet{Petrovich2020} showed that, even for $\psi_c \sim 0^\circ$, a resonance encountered as the young protoplanetary disk dissipates can excite an inner planet to high obliquities, even favoring a polar orbit given appropriate initial conditions. To summarize the model, consider a system with a close-in planet and a distant (few astronomical units) giant planet, like WASP-107, after the disk interior to the outer planet has been cleared but the disk exterior remains. The external gaseous disk induces a nodal precession of the outer planet at a rate proportional to the disk mass (Eq.~\ref{eq:nodal precession rate} with $b \mapsto c$ and $c \mapsto \text{disk}$). The outer planet still induces a nodal precession on the inner planet according to Eq.~\ref{eq:nodal precession rate}. If at first the rate $d\Omega_c/dt > d\Omega_b/dt$, then as the disk dissipates (and $M_\text{disk}$ decreases) the precession rate for planet c will decrease until it matches the precession rate of the inner planet. At this point the system will pass through a secular resonance, driving an instability which tilts the inner planet to a high obliquity; a small initial obliquity of a few degrees can quickly reach $90^\circ$. Additionally, depending on the relative strength of the stellar quadrupole moment and GR effects, the inner planet may obtain a high eccentricity (if GR is unimportant), a modest eccentricity (if GR is important), or a circular orbit (if GR dominates). Tidal forces can circularize the orbit, although the planet may retain a detectable eccentricity even after several gigayears. This process well explains the polar, close-in, and eccentric orbits of small planets like HAT-P-11b. Nodal precession alone is unable to explain the eccentricity of such planets.

Given the planet and stellar properties of the WASP-107 system, we calculated the instability criteria developed in \citet{Petrovich2020}. The steady-state evolution of the system can be inferred by comparing the relative strength of GR ($\eta_\text{GR}$) with the stellar quadrupole moment ($\eta_\star$). We found that $\eta_\text{GR} > \eta_\star + 6$ at 99.76\% confidence, $\eta_\star + 6 > \eta_\text{GR} > 4$ at 0.155\% confidence, and $\eta_\text{GR} < 4$ at 0.084\% confidence (i.e., $\eta_\text{GR} \sim 30 - 80$ and $\eta_\star \sim 1$). Thus WASP-107b is stable against eccentricity instabilities and lives in the polar, circular region of parameter space in Fig.~4 of \cite{Petrovich2020}.

We calculated the final obliquity of WASP-107b using the procedure outlined in \citet{Petrovich2020}, incorporating the uncertainties in $M\sin\iorbc$ and $P_c$ and integrating over all possible initial obliquities for the outer planet. Evaluating their Eq.~(3), we found that the resonance that drives the inner planet to high obliquities is always crossed. We calculated the adiabatic parameter $x_\text{ad} \equiv \tau_\text{disk} / \tau_\text{adia}$ from the disk dispersal timescale and the adiabatic time (their Eq.~7), taking $\tau_\text{disk}$ to be 1~Myr. In the orbital configurations where $x_\text{ad} > 1$ (adiabatic crossing) we computed the final obliquity from their Eq.~(12) ($I_\text{crit}$). Otherwise, the final obliquity was set to $I_\text{non-ad}$ from their Eq.~(15).

The resulting probability of the final obliquity of WASP-107b is 7.6\% for a nonpolar (but oblique) orbit and 92.4\% for a polar orbit. A polar orbit is likely if the outer planet's orbit is inclined at least $\sim 8^\circ$, and is guaranteed for $\psi_{\text{init},c} \gtrsim 25^\circ$. In an equivalent parameterization, \cite{Petrovich2020} explicitly predict a polar orbit for WASP-107b if the mass and semiminor axis of WASP-107c satisfy $(b_c / 2~\AU)^3 > (M_c / 0.5~\Mjup)$. Since we only have a constraint on $M\sin\iorbc$, this condition is satisfied if $\iorbc \in [60^\circ - 90^\circ]$. Such a viewing geometry, in conjunction with an obliquity of $\psi_c > 25^\circ$, is plausible given the likely stellar orientation (Section~\ref{sec:stellar inclination}).

A key deviation from this model is that while the orbit of WASP-107b is indeed close to polar, it is quite definitively retrograde. In the disk dispersal-driven tilting scenario, the inner planet approaches a $\psi = 90^\circ$ polar orbit from below and stops at $\psi_b = 90^\circ$. In order to reach a super-polar/retrograde orbit, %WASP-107b could have started with a significant obliquity at formation. For instance, a $\sim 10^\circ$ initial obliquity would lead to a $100^\circ$ final orbit (Cristibol Petrovich, private communication). Alternatively, 
WASP-107c must have a significant obliquity, either primordial from formation or through a scattering event~\citep{Petrovich2020}. As we alluded to in Section~\ref{sec:nodal precession}, a scattering event could also explain the moderate eccentricities of the outer giants WASP-107c and HAT-P-11c, and could easily give WASP-107c a high enough obliquity to guarantee a polar/super-polar configuration for WASP-107b~\citep{Huang2017}. % "We have discussed this possibility in our revised version of the paper, which can also respond your question 3 below on non-zero obliquities of the outer Jovians."
In fact a scattering event is more likley to produce the modest obliquity for planet c needed to produce a super-polar orbit under the disk dispersal framework than it is to produce the large ($\psi_c \gtrsim 40 - 50^\circ$) obliquity needed to excite either Kozai--Lidov or nodal precession cycles.

\section{Discussion and Conclusion}
\label{sec:conclusion}

We observed the RM effect during a transit of WASP-107b on 2020 February 26, from which we derived a near-polar and retrograde orbit as well as a low stellar $\vsini$. This low $\vsini$ implies that we are viewing the star close to one of its poles, reinforcing the near-polar orbital configuration of WASP-107b. However, we are unable to conclusively say how WASP-107b acquired such an orbit. Nodal precession or disk dispersal-driven tilting are both plausible mechanisms for producing a polar orbit, while Kozai--Lidov oscillations may be possible but only for a very narrow range of face-on orbital geometries for WASP-107c. RV observations \citep{Piaulet2020} as well as constraints on the velocity of the escaping atmosphere of WASP-107b~(e.g., \citealt{Allart2019}, \citealt{Kirk2020}, Spake, J. J. et al. 2020, in preparation) are consistent with a circular orbit. The eccentricity damping timescale due to tidal forces is only $\sim 60$~Myr~\citep{Piaulet2020}, so this is not unexpected. While a circular orbit does not rule out any of these pathways, only disk dispersal-driven tilting can explain both the eccentric and polar orbit of WASP-107b's doppelganger HAT-P-11 b.

Since all three scenarios depend on the obliquity of the outer giant planet, measuring the mutual inclination of planet b and c is essential to understand the dynamics of this system. This has been done for similar system architectures such as HAT-P-11~\citep{Xuan2020} and  $\pi$~Men~\citep{Xuan2020, DeRosa2020} by observing perturbations in the astrometric motion of the star due to the gravitational tugging of the distant giant planet, using data from Hipparcos and Gaia. Unfortunately WASP-107 is significantly fainter~~\citep[$V = 11.5$;][]{Anderson2017} and barely made the cutoff in the Tycho-2 catalog of Hipparcos \citep[90\% complete at V=11.5;][]{Tycho2}. %https://ui.adsabs.harvard.edu/abs/2000A&A...355L..27H
The poor Hipparcos astrometric precision, combined with the small angular scale of the orbit of WASP-107 on the sky (10 - 30 $\mu$as), prevents a detection of the outer planet using astrometry. Assuming future Gaia data releases have the same astrometric precision as in DR2 ($44~\mu$as for WASP-107), WASP-107c will be at the threshold of detectability using the full five-year astrometric time series.

On the population level, the disk dispersal-driven model favors low-mass and slowly rotating stars due to its dependence on the stellar quadrupole moment, and also can explain eccentric polar orbits. Since nodal precession has no stellar type preference nor a means of exciting eccentric orbits, measuring the obliquities and eccentricities for a population of close-in Neptunes will be essential for distinguishing which process is the dominant pathway to polar orbits. Additionally a large population is needed to determine if the overall distribution of planet obliquities is consistent with catching systems at different stages of nodal precession, or if there are indeed two distinct populations of aligned or polar close-in Neptunes. As these models all depend on the presence of an outer giant planet, long-baseline RV surveys will be instrumental for discovering the nature of any perturbing companions (e.g. Rosenthal et al. submitted). Moreover RV monitoring of systems with small planets that already have measured obliquities, but do not have mass constraints or detected outer companions, will further expand this population. Recent examples of such systems include
Kepler-408b~\citep{Kamiaka2019}, 
AU Mic b~\citep{Palle2020}, 
HD 63433 (b, \citealt{Mann2020}; and c, \citealt{Dai2020}), 
K2-25b~\citep{Stefansson2020},  
and DS Tuc b \citep{Montet2020, Zhou2020}.
Comparing the proportions of systems with and without companions which have inner aligned or misaligned planets will further illuminate the likelihood of these different dynamical scenarios.

\acknowledgements

We thank Konstantin Batygin, Cristobol Petrovich, and Jerry Xuan for helpful comments and productive discussions on orbital dynamics, and Josh Winn for constructive feedback that improved this manuscript. 
R.A.R.\ and A.C.\ acknowledge support from the National Science Foundation through the Graduate Research Fellowship Program (DGE 1745301, DGE 1842402). 
C.D.D.\ acknowledges the support of the Hellman Family Faculty Fund, the Alfred P.\ Sloan Foundation, the David \& Lucile Packard Foundation, and the National Aeronautics and Space Administration via the TESS Guest Investigator Program (80NSSC18K1583).  
I.J.M.C.\ acknowledges support from the NSF through grant AST-1824644. 
D.H. acknowledges support from the Alfred P. Sloan Foundation, the National Aeronautics and Space Administration (80NSSC18K1585, 80NSSC19K0379), and the National Science Foundation (AST-1717000). 
E.A.P.\ acknowledges the support of the Alfred P.\ Sloan Foundation. 
L.M.W.\ is supported by the Beatrice Watson Parrent Fellowship and NASA ADAP Grant 80NSSC19K0597.

We thank the time assignment committees of the University of California, the California Institute of Technology, NASA, and the University of Hawai`i for supporting the TESS--Keck Survey with observing time at the W. M. Keck Observatory.  
We gratefully acknowledge the efforts and dedication of the Keck Observatory staff for support of HIRES and remote observing.  We recognize and acknowledge the cultural role and reverence that the summit of Maunakea has within the indigenous Hawaiian community. We are deeply grateful to have the opportunity to conduct observations from this mountain.  
\facility{Keck I (HIRES)}

\software{
% RadVel \citep{radvel}, 
\texttt{emcee} \citep{emcee},
\texttt{corner.py} \citep{corner},
\texttt{REBOUND/REBOUND/x} \citep{rebound, {reboundx}},
\texttt{SciPy} \citep{scipy},
\texttt{NumPy} \citep{numpy},
\texttt{Matplotlib} \citep{matplotlib}
}

\bibliography{references}
\bibliographystyle{aasjournal}


\end{document}